# Drying-Induced Atomic Structural Rearrangements in Sodium-Based Calcium-Alumino-Silicate-Hydrate Gel and the Mitigating Effects of ZrO$_2$ Nanoparticles

Kengran Yang, V. Ongun Özçelik, Nishant Garg, Kai Gong and Claire E. White*

Department of Civil & Environmental Engineering

and

Andlinger Center for Energy and the Environment,

Princeton University, Princeton, USA

## Abstract

Conventional drying of colloidal materials and gels (including cement) can lead to detrimental effects due to the buildup of internal stresses as water evaporates from the nano/microscopic pores. However, the underlying nanoscopic alterations in these gel materials that are, in part, responsible for macroscopically-measured strain values, especially at low relative humidity, remain a topic of open debate in the literature. In this study, sodium-based calcium-alumino-silicate-hydrate (C-(N)-A-S-H) gel, the major binding phase of silicate-activated blast furnace slag (one type of low-CO$_2$ cement), is investigated from a drying perspective, since it is known to suffer extensively from drying-induced microcracking. By employing *in situ* synchrotron X-ray total scattering measurements and pair distribution function (PDF) analysis we show that the significant contributing factor to the strain development in this material at extremely low relative humidity (0%) is the local atomic structural rearrangement of the C-(N)-A-S-H gel, including collapse of interlayer spacing and slight disintegration of the gel. Moreover, analysis of the medium range (1.0 – 2.2 nm) ordering in the PDF data reveals that the PDF-derived strain values are in much closer agreement (same order of magnitude) with the macroscopically measured strain data, compared to previous results based on reciprocal space X-ray diffraction data. From a mitigation standpoint, we show that small amounts of ZrO$_2$ nanoparticles are able to actively reinforce the structure of silicate-activated slag during drying, preventing atomic level strains from developing. Mechanistically, these nanoparticles induce growth of a silica-rich gel during drying, which, via density functional theory calculations, we show is attributed to the high surface reactivity of tetragonal ZrO$_2$.

# Introduction

Drying is a common (and potentially detrimental) processing technique used in a variety of industries, including sol-gel synthesis and ceramics. The removal of fluid (usually water) from a porous material can lead to a buildup of internal stresses.[1] Typically the body will not dry uniformly, and therefore a drying front will develop in the material where the dried outer component wants to shrink, whereas the moist interior restrains shrinkage. This mismatch in strain leads to tensile stresses developing on the surface of the material, and if these stresses exceed the tensile strength, the material will crack. Mitigation strategies include careful control of material thickness,[2] use of surface coatings to prevent cracking during evaporation[3] and controlled drying.[4] One class of material that is used in significant quantities around the world that is prone to suffer from drying, and therefore cracking, is concrete. Although there are several viable mediating technologies available for Portland cement-based concrete to prevent drying-induced microcracking, the underlying mechanisms responsible for the macroscopically-measured strains remain somewhat unclear, especially those occurring at low relative humidity (RH, < 40%). Moreover, the development of low-$CO_2$ cement alternatives, such as alkali-activated materials (AAMs), has led to a reexamination of cement degradation phenomena, including drying-induced microcracking.

Given that ordinary Portland cement (OPC) production accounts for 5 – 8% of anthropogenic $CO_2$ emissions,[5] there is a pressing need to develop and implement sustainable alternatives. AAMs are one of the most competitive alternatives, and have been shown to emit less $CO_2$ (~ 40 – 80%) compared with OPC.[6-8] AAMs utilize aluminosilicate-rich precursor materials, including industrial by-products such as ground granulated blast-furnace slag, fly ash from coal-fired power plants and calcined clays (e.g., metakaolin), which form mechanically hard binders (gels) when activated by alkaline solutions (or solids, as is the case for 1-part mixes[9,10]).[6,11] Furthermore, given correct mix designs, AAMs have comparable mechanical performance[12] and cost[7] to OPC, and can be tuned to have superior properties via specific chemical compositions, such as high thermal performance[13] and low permeability.[14]

Nevertheless, questions remain regarding the long-term durability of AAMs, with the underlying degradation mechanisms often founded at the atomic/nanoscale, such as carbonation-induced chemical reactions[15-18] (from atmospheric and accelerated $CO_2$ conditions) and sulfate attack of the binder gel.[19,20] Progress is being made to elucidate the mechanisms responsible for chemical degradation of different types of AAMs, with the aim to pinpoint which mix designs are most resistant to different forms of degradation.[16,21,22] However, microcracking remains an outstanding durability and aesthetic issue for silicate-activated slag-based AAMs, and since silicate-activated slag possesses superior setting times,[23] strength development,[23] and permeability[14] compared with other AAMs, having a fundamental scientific understanding of the underlying cause(s) of microcracking is needed together with potential solutions for minimizing/mitigating this issue.

The susceptibility of silicate-activated slag to microcracking has been known for decades,[24-27] where the extent of microcracking in silicate-activated slag paste is considerably larger compared with OPC paste.[27] Collins and Sanjayan proposed that the difference in drying shrinkage between silicate-activated slag and OPC is due to their different pore size distributions, imposing different levels of capillary force on the OPC/silicate-activated slag paste as water is removed from the pores.[27] However, it is known that the pretreatment process (i.e., removal of pore liquid) required by the two major pore characterization techniques (i.e., nitrogen sorption and mercury intrusion porosimetry) may alter the pore structure of cementitious materials to a non-negligible extent.[28-33] More recently, based on the permeability results from the beam-bending technique, Scherer *et al.* estimated the pore diameters of OPC paste to be ~ 1.5 – 5 nm,[28] which was also observed by Zeng *et al.* using nitrogen sorption with freeze-drying as the method of pretreatment.[34] Recent nitrogen sorption measurements on silicate-activated slag performed by Blyth *et al.*[14] with pretreatment via isopropanol solvent exchange showed that the major pore size obtained from the desorption curve of the isotherm is ~ 3 – 4 nm, which is close to that of OPC. However, it was also shown by Blyth *et al.* that the permeability of silicate-activated slag is lower than that of OPC paste (0.0001 $nm^2$ compared with 0.005 $nm^2$ at 7 days for water/precursor ratio of 0.5 (w/c ratio for OPC)).[14] As reported by Scherer, the permeability has an inverse relationship to the stresses that develop during drying,[35] and therefore, this difference in permeability between silicate-activated slag and OPC paste may be responsible in part for the susceptibility of silicate-activated slag paste to microcrack.

During drying of cementitious materials, it is generally accepted that there are no significant changes occurring to the atomic structure of the paste. Internal stresses lead to shrinkage of the body via consolidation of the particles (i.e., rearrangement of the cement grains). Only when the RH reaches very low levels (< 20%) does water evaporate from the interlayer spacing of calcium-silicate-hydrate (C-S-H) gel, the major binding phase of OPC paste, leading to a collapse of this interlayer spacing and an associated additional shrinkage at the macroscopic level.[36] However, a few articles have pointed to additional changes in the C-S-H gel structure due to drying, specifically intra- and inter-granular cohesion of C-S-H,[37] changes in the polymerization state of the silicate species[38] and a decrease in atomic ordering (as measured via the width of $^{29}$Si nuclear magnetic resonance (NMR) peaks, with the proposed formation of new Ca-O-Si bonds as water is removed from the interlayer).[39] Additional research is required to substantiate these results, and to uncover the link between these proposed changes and the macroscopically measured drying shrinkage.

One experimental technique that is capable of probing the local atomic structure of disordered/amorphous materials is pair distribution function (PDF) analysis. Using this technique, we recently found that the atomic structure of sodium-containing calcium-alumino-silicate-hydrate (C-(N)-A-S-H) gel in silicate-activated slag is more disordered (i.e., more amorphous) than synthetic C-S-H gel, indicating that C-(N)-A-S-H gel may be thermodynamically less stable.[40] Based on this observation, we hypothesize that the atomic structure of C-(N)-A-S-H gel in silicate-activated slag may undergo atomic structural changes (rearrangements) during drying, which contribute to the macroscopically measured drying shrinkage at low RH values.

 Here, we directly measure changes in the local atomic structure of silicate-activated slag *in situ* as the material is subjected to drying conditions using synchrotron-based X-ray PDF analysis. By tracking the evolution of specific atom-atom correlations we are able to quantify the extent of change to the atomic structure as a function of RH, and also pin-point which atom-atom correlations are affected by drying. Furthermore, we show it is possible to directly measure drying-induced strain at the nanoscale via shifts of the higher $r$ atom-atom correlations (between 10 and 22 Å), providing new insight on the susceptibility of C-(N)-A-S-H gel to undergo alterations during drying.

To control/limit changes at the nanoscale in C-(N)-A-S-H gel during drying, we have investigated the impact of zirconia ($ZrO_2$) nanoparticles on the structure and stability of C-(N)-A-S-H gel, whose catalytic effect has been well documented.[41,42] Due to the high surface-to-volume ratio, nanoparticles may potentially improve the performance of cementitious materials by acting as extra nucleation sites for gel growth or providing a filler effect. Numerous studies have been conducted on adding nanoparticles to cementitious materials (albeit sometimes at extremely high concentrations (> 1% wt.) which may not be economically viable),  revealing that the addition of nanoparticles can accelerate the reaction process,[25,43-47] increase strength,[25,44,48,49] and reduce porosity.[25,48] However, there is limited literature available on the impact of nanoparticles on cementitious materials under drying conditions. Yang *et al*. reported that adding nano-$TiO_2$ in silicate-activated slag reduces the extent of drying shrinkage.[25] On the other hand, a review by Rashad showed that the addition of nano-silica to a cement-based mortar partially substituted by rice husk ash produced mix results for the extent of drying shrinkage.[50]

In this investigation, the impact of nano-$ZrO_2$ (stable under high pH conditions) on the reaction kinetics of silicate-activated slag has been elucidated using *in situ* X-ray PDF analysis, together with the influence of these nanoparticles on the drying-induced changes that occur when silicate-activated slag is exposed to low RH environments. Given that the nanoparticles are seen to drastically alter the nanoscale behavior of the material during drying, density functional theory (DFT) calculations have been used to uncover the mechanism by which the nanoparticle surfaces alter the evolution of the material during evaporation of the pore solution. The results provide new insight on the viability of using nano-$ZrO_2$ to limit the extent of drying-induced shrinkage strains in silicate-activated slag, and therefore mitigate the extent of microcracking typically seen in these sustainable cements.

# Materials and methods

## Material synthesis

Silicate-activated slag was synthesized using ground granulated blast-furnace slag (denoted as slag), with the slag composition (GranCem, Holcim) shown in Table 1. The slag was activated using a sodium silicate solution with a $Na_2O$ wt. % of 7 (i.e., 7g of $Na_2O$ per 100g of slag), since preliminary experiments showed that drying-induced microcracking is prevalent at this alkali concentration. The sodium silicate solution was synthesized by dissolving anhydrous sodium metasilicate ($Na_2SiO_3$, reagent grade, Sigma-Aldrich) in deionized water. To ensure that the silicate species (i.e., oligomers) in the solution reached equilibrium, the sodium silicate solution was mixed for at least 24 hrs using a magnetic stirrer bar. The water-to-slag wt. ratio was set at 0.44 for all samples in order to maintain good workability of the paste. The pastes were synthesized by manual mixing for ~ 2 min until the samples appeared homogeneous. For the silicate-activated slag paste containing nano-$ZrO_2$, 0.167 wt. % (by mass of anhydrous slag) of $ZrO_2$ nanoparticles (supplied as a 10% wt. dispersion in $H_2O$, Sigma-Aldrich) was added to the sodium silicate solution and stirred thoroughly before the addition of slag powder. The maximum size of the $ZrO_2$ nanoparticles was less than 100 nm, as reported by the manufacturer.

After mixing, control samples (not exposed to drying conditions, one with and one without nano-$ZrO_2$) were prepared by suctioning the paste into 1mm diameter polyimide capillary tubes using a syringe. The two ends of the capillary were then sealed using quick-setting epoxy. These control samples were measured at the start and end of the *in situ* drying measurements to assess if the alkali-activation reaction contributed to the local atomic changes seen in the samples exposed to drying conditions. The remaining pastes were sealed and left to cure for 24 hrs, after which the samples were ground into fine powders using a mortar and pestle in a 96% RH glove bag (using a $K_2SO_4$ saturated salt solution and $N_2$ gas), and then immediately loaded into polyimide capillaries and sealed using porous glass wool in order to enable the nitrogen gas with different RH to flow through the powdered samples during the *in situ* PDF measurements.

The RH of the gas used during the *in situ* PDF measurements was controlled by flowing dry $N_2$ gas through bubblers filled with different saturated salt solutions. Three different RH values were used: 0% (using dry $N_2$ gas), 43% and 96%. The 43% and 96% RH conditions were attained by using supersaturated $K_2CO_3$ and $K_2SO_4$ solutions, respectively (both from Sigma-Aldrich, reagent grade).

Table 1. Oxide Composition (wt. %) of the slag used in this investigation. From ref. 15.

| CaO | SiO$_2$ | Al$_2$O$_3$ | MgO | SO$_3$ |
|------|------|------|------|------|
| 42.5 | 34.5 | 11.7 | 7.3 | 1.7 |

## X-ray data collection and analysis

The X-ray total scattering experiments were conducted at the Advanced Photon Source, Argonne National Laboratory on the 11-ID-B beamline. The silicate-activated slag samples were mounted and aligned in the gas cell[51] under ambient conditions. Each sample was analyzed using a wavelength of 0.2112 Å and a two-dimensional image plate detector.[52] The detector-to-sample distance was ~ 175 mm. For the samples exposed to drying conditions, a bubbler containing mineral oil was connected to the gas cell arrangement to check for adequate gas flow through the sample, and to indicate if a blockage had occurred. A pre-scan was conducted for each sample for 10 min, where humid $N_2$ gas with 96% RH flowed through the sample. Afterwards, the gas was switched to 0% or 43% RH, and data were acquired every 2.5 min. The duration of each test was determined by screening data until no significant changes in the PDFs were observed. The control sample was measured on the gas cell before and after each *in situ* drying measurement. Data conversion from 2D to 1D was carried out using the program Fit2D with $CeO_2$ as the calibration material.[53,54] The PDF, G(r), was obtained by taking a sine Fourier transform of the measured total scattering function, S(Q), as shown in eqn. 1, where Q is the momentum transfer given in eqn. 2.[55]

$$G(r) = \frac{2}{\pi}\int_{Q_{min}}^{Q_{max}} Q[S(Q)\text{-}1]\sin{(Qr)}dQ$$

$$\text{(1)}$$

$$Q = \frac{4\pi sin\theta}{\lambda}$$

$$\text{(2)}$$

Standard data reduction procedures were followed to obtain the PDF using PDFgetX2,[56] with a $Q_{max}$ of 22 Å$^{-1}$.

The control silicate-activated slag samples (with and without nanoparticles) were also measured on 11-ID-B approximately five months after synthesis, in order to investigate the atomic changes due to the alkali-activation reaction. Due to slight differences in the beamline setup for the two experiments, a normalization procedure with respect to the photon count has been carried out for all data presented in this investigation.

## Density functional theory calculations

DFT calculations were performed to compare the interaction energies of silicate ions (originating from the pore solution in the experiments) with different solid surfaces (zirconia vs. C-(N)-A-S-H present in the samples). These calculations were performed using the generalized gradient approximation (GGA) including van der Waals corrections.[57] The projector-augmented wave potentials (PAW) were used,[58] and the exchange-correlation potential was approximated with the Perdew-Burke-Ernzerhof functional.[59] The Brillouin zone was sampled using 5×5×3 *k*-points in the Monkhorst-Pack scheme where the convergence in energy as a function of the number of *k*- points was tested. The energy convergence value between two consecutive steps was chosen as $10^{-4}$ eV with an energy cutoff value of 500 eV. A maximum force of 0.05 eV/Å was allowed on each atom. The DFT calculations were carried out using the VASP software.[60]

# Results & discussion

## Impact of drying on C-(N)-A-S-H gel

### Alterations to the local atomic structure

Figure 1 shows the PDF curves for the silicate-activated slag sample exposed to extreme drying conditions (0% RH, in Figure 1a) together with the data for the control sample (Figure 1b). Peak assignments given in Figure 1 are explained in detail in the Supporting Information, where additional information on the local atomic structural changes seen during the alkali-activation reaction is provided. The PDF curves in Figure 1a have been normalized with respect to the maximum T-O peak intensity (T represents Si or Al) of the initial PDF data set at ~ 1.65 Å, since this nearest-neighbor bonding environment, specifically the number of tetrahedral Si/Al units in the sample, should be relatively constant throughout the drying process. A similar normalization process for PDF data has been used in our previous study.[61] The data from the control sample are not normalized because the experimental setup at 3624 h is slightly different from the rest, which may introduce artifacts in PDF peak intensities if normalization was carried out. The 0 hr data set in Figure 1a is defined as the start of 0% RH gas flow, at which point the silicate-activated slag control sample shown in Figure 1b has been curing for 24 hrs.

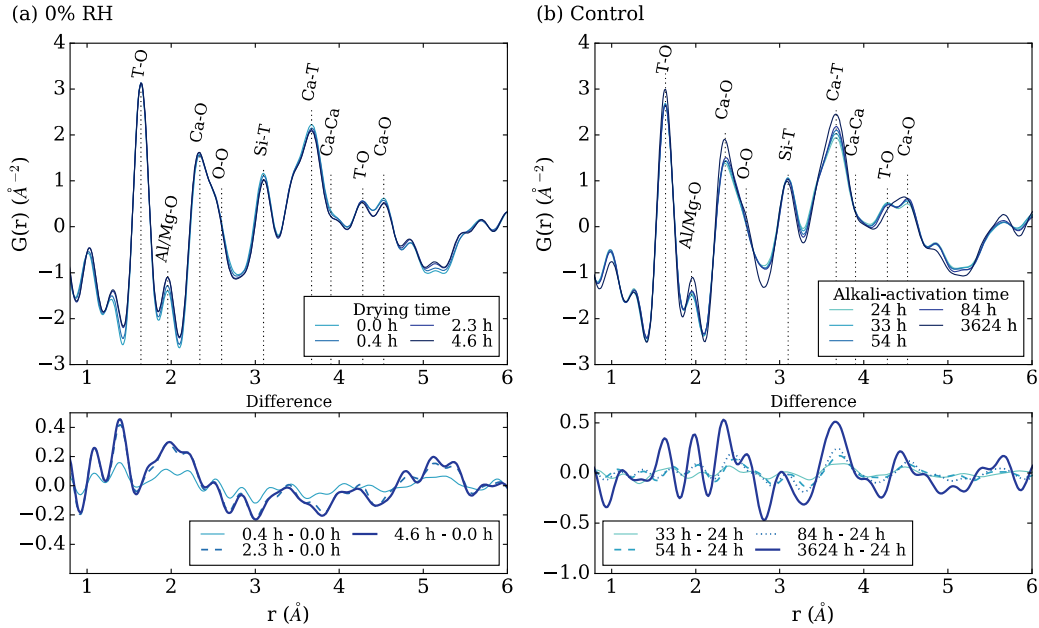

Figure 1. Synchrotron X-ray PDF curves of silicate-activated slag subjected to 0% RH (a), and during the alkali-activation reaction (b, denoted as "control"). PDF curves at different times during drying/alkali-activation (top), and difference curves of the PDF data obtained via subtraction of the initial PDF data set (bottom).

It is clear from Figure 1 that the atomic structural changes that occur due to drying (Figure 1a) are significantly different from the changes seen as a result of the alkali-activation reaction (Figure 1b). The Ca-T correlation at ~3.65 Å, which provides an indication of the amount of C-(N)-A-S-H gel in the sample,[15] is seen to decrease slightly with time in Figure 1a, which is opposite to the trend in Figure 1b where the gel continues to precipitate throughout the course of the measurement. The decrease of the Ca-Ca (at 3.90 Å) and Ca-O (at 4.53 Å) peak intensities in Figure 1a implies that changes are occurring to the CaO layers (intra- or inter-layer Ca environments) of the C-(N)-A-S-H gel. The Si-T correlation is observed to decrease in intensity in Figure 1a, in contrast to Figure 1b where this peak does not change significantly during the alkali-activation reaction.

Overall, the stark differences between Figure 1a and 1b show that there are observable changes occurring in the atomic structure of silicate-activated slag during drying at 0% RH. The C-(N)-A-S-H gel is experiencing rearrangements at the atomic length scale, where the Ca-O-T linkages are being broken together with a decrease in the mean chain length (MCL) of the (alumino)silica units or a reduction in the extent of cross-linking (seen via the loss of Si-O-T linkages). A decrease in the MCL has also been observed in an OPC system with 22% wt. of slag and 22% wt. of silica fume as supplementary cementitious materials by [29]Si NMR spectroscopy when subjected to drying.[62] It is likely that as the interlayer water is being pulled out (which occurs after the evaporation of the water from the gel pores), the large drying-induced capillary stresses being exerted on the gel structure are high enough to cause bond breakage and subsequent rearrangement at the atomic length scale. Hence, these data show that the atomic structure of C-(N)-A-S-H gel likely undergoes disintegration to a certain degree as a result of exposure to extreme drying conditions. Furthermore, by comparing the difference curve in Figure 1a with PDF data in the literature on liquid water,[63,64] it is clear that the changes in the PDF curves of the silicate-activated slag sample undergoing drying are not only due to the removal of water molecules, but are also attributed to structural rearrangements of the atomic structure of silicate-activated slag (for further discussion see Supporting Information).

It has been recently reported in the literature that C-S-H gel in white Portland cement paste undergoes nanoscale changes as a result of drying, where the small-angle X-ray scattering data were fit using a disk-shaped model, and the data revealed that at low RH the disks reduce in thickness and width, which is in agreement with the slight disintegration-like behavior seen in Figure 1a.[65] On the other hand, the PDF curves for the silicate-activated slag sample exposed to moderate drying conditions (43% RH) do not show obvious changes in the atomic structure (see Figure S2a in Supporting Information), and therefore disintegration of the C-(N)-A-S-H gel is only realized at extreme drying conditions.

**Kinetics of local structural changes**

To quantify the rate of change in the atomic structure of C-(N)-A-S-H gel as a result of drying (in both the 0% and 43% RH environments), two ratios have been calculated from peak intensities in the PDF data. It has been previously shown that the degree of polymerization of the C-(N)-A-S-H gel can be inferred from the maximum intensity of the Si-T correlation at ~ 3.1 Å divided by the intensity of the T-O peak at ~ 1.65 Å (i.e., (Si-T)÷(T-O)).[66] Likewise, information on the quantity of Ca-T linkages in the gel can be determined via (Ca-T)÷(T-O). The reason for dividing through by the T-O peak intensity is to account for any sample density increase/decrease that may occur due to the gas stream pushing powder in/out of the X-ray beam.

The (Si-T)÷(T-O) and (Ca-T)÷(T-O) ratios are given in Figure 2 for the samples exposed to 0 and 43% RH, together with these ratios for the control sample as the alkali-activation reaction evolves. After drying commences in the sample exposed to a 0% RH nitrogen gas flow, the (Si-T)÷(T-O) ratio is seen to continually decrease until equilibrium is reached after ~ 4 hrs. In contrast, the (Si-T)÷(T-O) ratios for the control sample and the sample exposed to 43% RH stay relatively stable during the measurement time frame (up to 9 hrs). Furthermore, the (Ca-T)÷(T-O) ratio for the sample exposed to the 0% RH environment also shows a decreasing trend that is very similar to the (Si-T)÷(T-O) ratio trend. Hence, these data in Figure 2 provide quantitative evidence of the rearrangement of C-(N)-A-S-H gel, where breakage of the C-(N)-A-S-H gel structure involves segmentation along its elongated axis, where the (alumino)silica chains and CaO layers are seen to rupture simultaneously (Figure 3 shows this process schematically).

Similar segmentation of a elongated C-S-H unit (~ 35 nm diameter, modeled as a disk shape) into shorter ones (~ 10 nm) below a RH of 40% was also identified in white Portland cement,[65] as mentioned earlier in this article. Moreover, we showed in a previous study that the nanoscale morphology of freeze-dried silicate-activated slag has a globular-like appearance as measured using helium ion microscopy, where the globules are of sizes ranging from ~10 to 100nm.[67] Although these globules are larger than the disk-like particle sizes obtained using small-angle scattering,[65] the helium ion microscopy data supports our hypothesis that the C-(N)-A-S-H gel undergoes a certain degree of disintegration due to exposure to extreme drying conditions.

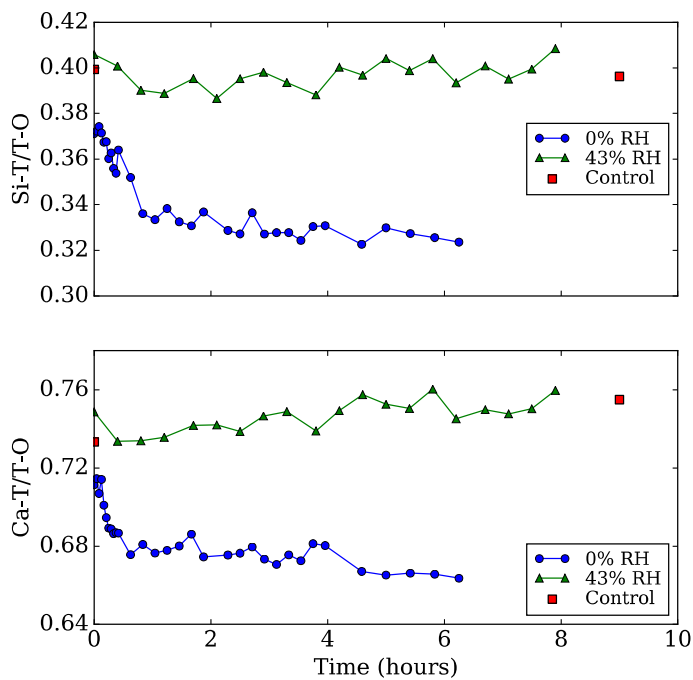

Figure 2. Evolution of normalized Si-T (at ~3.1 Å) and Ca-T (at ~3.65 Å) peak intensities of silicate-activated slag at 0% RH, 43% RH and in the control (sealed) environment, all normalized with respect to the T-O peak intensity at ~1.65 Å.

Based on the Kelvin equation, Jennings proposed that different mechanisms of drying operate at different levels of RH.[68] At a RH of 0%, the interlayer spacing of the C-(N)-A-S-H gel will be emptied, while pores of size 2 – 5 nm will be emptied at a RH of 40%.[68] In fact, a decrease in basal (interlayer) spacing of the C-(N)-A-S-H atomic structure is observed in the reciprocal-space diffraction data of the silicate-activated slag sample in the 0% RH environment (see Figure S4a in the Supporting Information), which has also been observed in C-S-H gel during drying using diffraction[69,70] and molecular simulations.[36] However, the extent of change in the interlayer spacing of the C-S-H gel is much larger (~ 2 Å)[69,70] than the amount measured in this study (0.6 Å, see Supporting Information). This is probably due to the difference in drying time, where our sample was dried for only a few hours, whereas the synthetic C-S-H samples were dried for a few weeks. The large difference in the amount of weight loss (more than 50% of the dried mass in the study by Gutteridge and Parrott[69] versus ~15% in our study (see Supporting Information)) also supports this explanation.

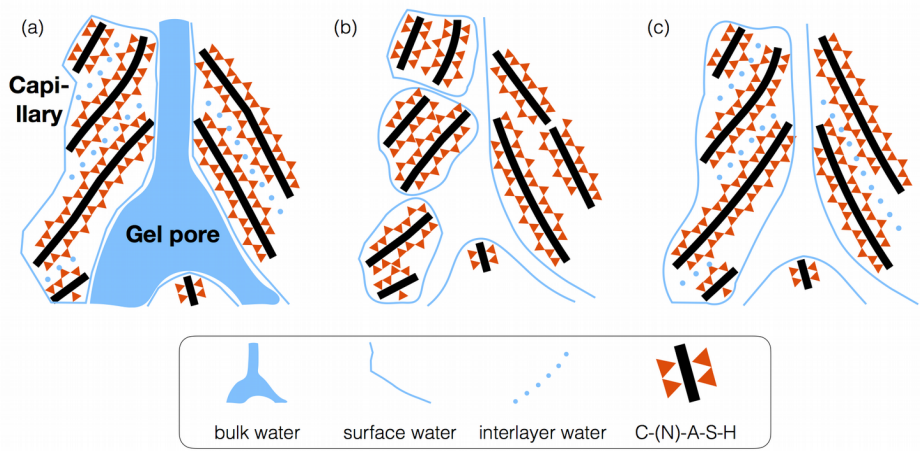

Figure 3. Schematic diagrams of C-(N)-A-S-H gel subjected to different RH at the nanoscale (~ 10 nm). (a) In the control environment, and (b) at 0% RH where all three types of water are removed (large C-(N)-A-S-H gel disintegrates into smaller ones, and its interlayer spacing decreases). (c) At 43% RH where only water in the gel pores is removed. The gel pores become smaller at smaller RH due to large capillary stresses being exerted on the material.

**Strain at the nanoscale**

The mid-range atomic ordering (10 – 22 Å) in the PDF data, as shown in Figure 4, contains information on the impact of drying on the interlayer collapse and associated development of nanoscale strain in silicate-activated slag. Peak shifts at various $r$ locations are found in the silicate-activated slag subjected to the 0% RH drying condition, as indicated by the black arrows in Figure 4a. However, shifts in the corresponding peaks in the silicate-activated slag subjected to 43% RH are minimal, as shown in Figure 4b. Quantification of this observation is given in Figure 5, where a summary of the extent of peak shifts is provided.

(a)

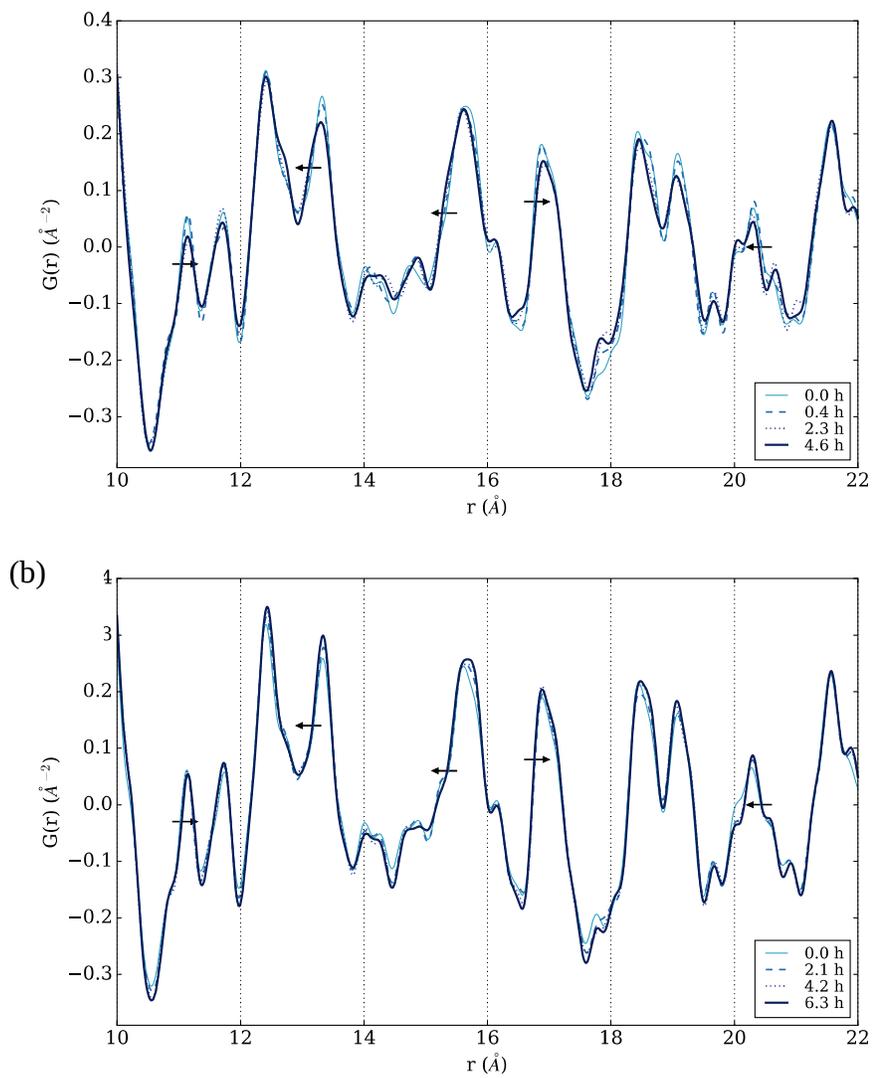

Figure 4. Synchrotron X-ray PDF curves of silicate-activated slag at (a) 0% RH and (b) 43% RH for different drying times, over an *r* range of 10 < *r* < 22 Å. Black arrows indicate the direction of peak shifts for the case of 0% RH.

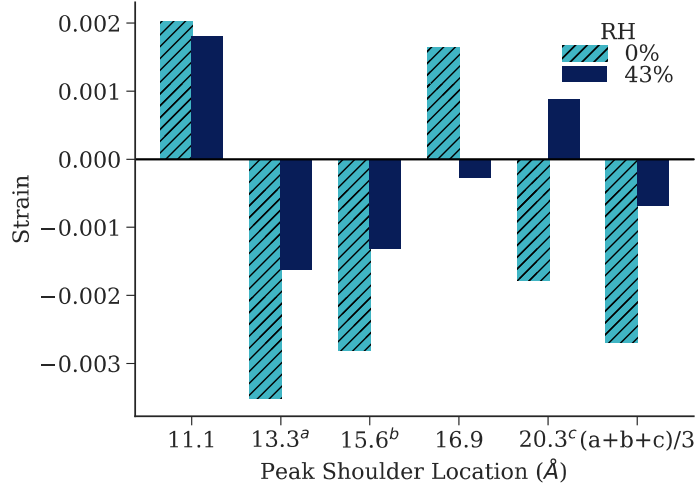

Figure 5. Quantification of the peak shifts in the PDF data for silicate-activated slag (Figure 4), given as strain values. A positive strain value indicates a shift of the peak shoulder towards larger atomic distances.

The extent of peak shift in the PDF data is represented using strain, defined as $(r_{end\ time} - r_{0\ h})/r_{0\ h}$, where $r_{0\ h}$ is the $r$ position before drying and $r_{end\ time}$ is the $r$ position after drying (i.e., $end\ time = 4.6\ h$ for 0% RH and $end\ time = 6.3\ h$ for 43% RH). For each peak, three intensity values (two for the peak at 20.3 Å because of its limited peak height) are selected along its shoulder, and the corresponding $r$ values (atomic distances) are recorded, with the average value reported in Figure 5. It is clear from this figure that the magnitude of peak shift for the silicate-activated slag in the 0% RH environment is larger than the 43% RH environment for all five selected peaks.

It is anticipated that a collapse of the interlayer spacing in the C-(N)-A-S-H gel will be apparent in the PDF data over an $r$ range of $10 - 22$ Å, assuming an initial interlayer spacing of ~ $11 - 14$ Å. The peaks at 13.3, 15.6 and 20.3 Å in Figures 4 and 5 do decrease in atomic distances, however, the 11.1 and 16.9 Å peaks are seen to increase continually as drying progresses. Given that the PDF data contain all atom-atom correlations in the material, the changes captured in Figures 4 and 5 will include not only quantitative information on the collapse of the interlayer spacing, but also all other atomic structural rearrangements that occur during drying (such as those associated with the disintegration as mentioned earlier, albeit manifested at a longer length scale). However, teasing out the individual atom-atom correlations at this length scale ($10 - 22$ Å) without an accurate structural model of C-(N)-A-S-H gel is extremely difficult.

The strain values presented in Figure 5 are within an order of magnitude of those measured at the macroscopic level for pure C-S-H gel (0.03),[36,69] OPC paste (0.0005 to 0.003)[71,72] and previous studies on silicate-activated slag (0.006)[25] and hydroxide-activated slag pastes (0.017).[70] Interestingly, the excessively large strain values reported from XRD-measured changes in the basal spacing for C-S-H gel (0.2),[36,69] which we also partially observe in the reciprocal space data in the Supporting Information for silicate-activated slag (0.6Å/14Å = 0.04) are not captured by the PDF data, and therefore the PDF peak shifts better represent the magnitude of nanoscopic shrinkage that occurs within the sample. The data presented in Figures 2 and 4b show that at 43% RH there are limited changes occurring to the atomic structure of the C-(N)-A-S-H gel, and therefore the corresponding strain values (Figure 5) are minimal (almost an order of magnitude smaller than those measured for 0% RH).

Many efforts have been made to link drying shrinkage of OPC observed at the macroscopic level with the underlying mechanisms originating at micro- and nanoscale. During the 1990's, Jennings and Xi identified knowledge gaps between the macroscopically measure drying strains and the underlying (and unresolved) multiscale mechanisms, and proposed a microscale model to help bridge this gap.[73,74] They used environmental scanning electron microscopy to provide experimental data for their model at the micron length scale, however, the strain measured using this technique was found to be 10 times larger than the macroscopic strain values.[73] Later, additional insight was provided by Thomas *et al.* using small-angle neutron scattering (SANS), where changes in the packing density and surface area of OPC were observed for different RH conditions.[75] They also noted that the SANS data was not able to elucidate information at the atomic level due to the resolution limit (in $Q$ space) of the instrument.

Recently, Pinson *et al.* modeled the relationship between drying shrinkage and RH for OPC, where they attributed the macroscopic length change to three factors: Laplace pressure, surface energy and loss of interlayer water.[36] For the strain induced by the loss of interlayer water they assumed a simple linear relationship between the interlayer spacing and contributions to the macroscopic strain, where a scaling factor of 0.1 was applied to the extent of change of the basal spacing. This was carried out in light of the experimental results of Gutteridge and Parrott[69] and Neubauer *et al.*,[76] and Pinson *et al.* attributed such large differences in the nanoscopic and macroscopic strain values to the possibility that only part of the nanoscopic strain is accommodated by the porous material and only some of the interlayer spaces undergo length changes. However, it is important to note that the basal spacing for C-S-H gel provides an average representation of interlayer spacing, and therefore if only part of the C-S-H underwent these changes, then the Bragg peak would not shift, but instead would broaden towards small *d* values (i.e., higher 2θ values). In fact, the XRD data provided in ref. 69 shows some signs of peak broadening, where a residual shoulder is present for the uncollapsed C-S-H interlayer spacing after drying. Hence, these data highlight the limitations of obtaining nanoscopic strain information from the shift of a single Bragg peak.

Hence, shrinkage at the nanoscale is a complex process consisting of multiple changes to the atomic structure of the gel (i.e., collapse of the interlayer spacing along with local structural rearrangements and changes in silica connectivity), and therefore the extent of shrinkage is more aptly captured from analysis of the PDF data (Figure 4) compared to conventional XRD analysis. Moreover, Pinson *et al.* showed that the maximum macroscopic shrinkage (at low RH) tends to be dominated by strain at the nanoscale (defined as "interlayer" in the article), and our data and subsequent analysis outlined above (Figures 4 and 5) provide more accurate estimates of the magnitude of this shrinkage (0.0027 for C-(N)-A-S-H gel at 0% RH, ~0.0028 for Pinson *et al.* obtained from Figure 6 in ref. 36) without any artificial manipulation of the strain value obtained from X-ray scattering results. Recently, such method of measuring nanoscopic strain from PDF data was used to study the behavior of C-S-H gel under compressive stress, where good agreement between nanoscopic and macroscopic strain was observed.[77] This further demonstrates the suitability of PDF data to measure nanoscopic strain.

### Effect of nano-ZrO$_2$ on drying behavior of C-(N)-A-S-H gel

The susceptibility of cementitious materials to chemical degradation and other stability issues (such as drying-induced shrinkage) has led to researchers experimenting with certain types of nanoparticles with the aim of improving cement performance. For instance, inert nanoparticles have been investigated for their ability to manipulate precipitation and nucleation of the main binder gel during hydration of OPC[44,46] or alkali-activation.[25,45] Here, nano-ZrO$_2$ is utilized as a potential method to augment the drying-induced atomic structural changes, where it is seen to drastically alter the behavior of silicate-activated slag at an RH of 0% (comparing Figure 6a and Figure 1a). The Ca-T peak in the nano-ZrO$_2$ sample increases continuously with time as drying progresses, as do the first and second nearest-neighbor Ca-O peaks, which is opposite to the changes seen to occur for the sample without nano-ZrO$_2$. Unexpectedly, the peak changes in the nano-ZrO$_2$ sample subjected to drying (Figure 6a) have some resemblance with the difference pattern shown in the silicate-activated slag sample during alkali-activation (Figure 1b and 6b), which is indicative of gel growth. In fact, direct comparison of these difference curves, as given in Figure 7, reveals that there are significant similarities in the two gels (the C-(N)-A-S-H gel that forms during alkali-activation and the "unconventional" gel that precipitates during drying in the sample containing nano-ZrO$_2$), especially over an *r* range of ~ 2 – 5 Å. On the other hand, the impact of nano-ZrO$_2$ on alkali-activation of the silicate-activated slag is minimal, where a comparison of the subtle difference of the evolution of the Ca-T peak can be found in the Supporting Information (Figure S7). Furthermore, as shown in the Supporting Information (Figure S9), the changes seen in the PDF data in Figure 6a are attributed to the atomic structural changes occurring in silicate-activated slag as opposed to separate atom-atom contributions coming from the nano-ZrO$_2$.

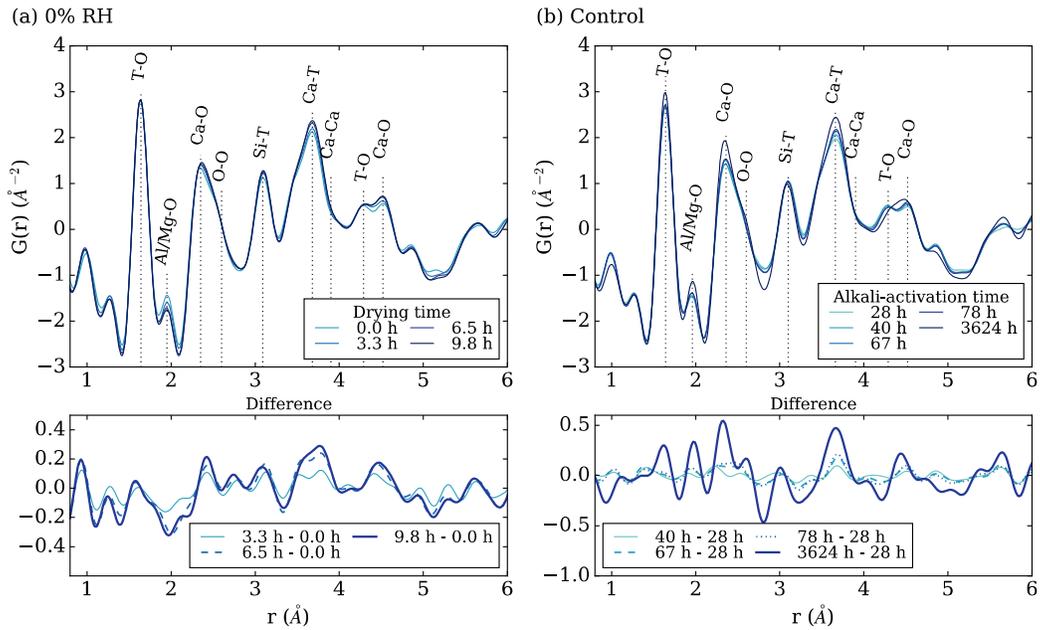

Figure 6. Synchrotron X-ray PDF curves of silicate-activated slag with nano-$ZrO_2$ subjected to 0% RH (a), and during the alkali-activation reaction (b, denoted as "control"). PDF curves at different times during drying/alkali-activation (top), and difference curves of the PDF data obtained via subtraction of the initial PDF data set (bottom).

It is observed in Figure 6a that the Si-T peak increases, which is not the case for silicate-activated slag samples undergoing alkali-activation (Figures 1b and S1). This atom-atom correlation denotes the Si-O-T linkages in the (alumino)silica chains, and therefore, an increase in this peak implies that more Si-O-T linkages form, leading to a higher degree of polymerization for the precipitated "unconventional" gel. A similar behavior is seen to occur in the nano-$ZrO_2$ sample exposed to 43% RH nitrogen gas flow, where the Ca-O (first and second nearest-neighbor), Ca-T and Si-T peaks increase as drying progresses (Figure S2b in the Supporting Information).

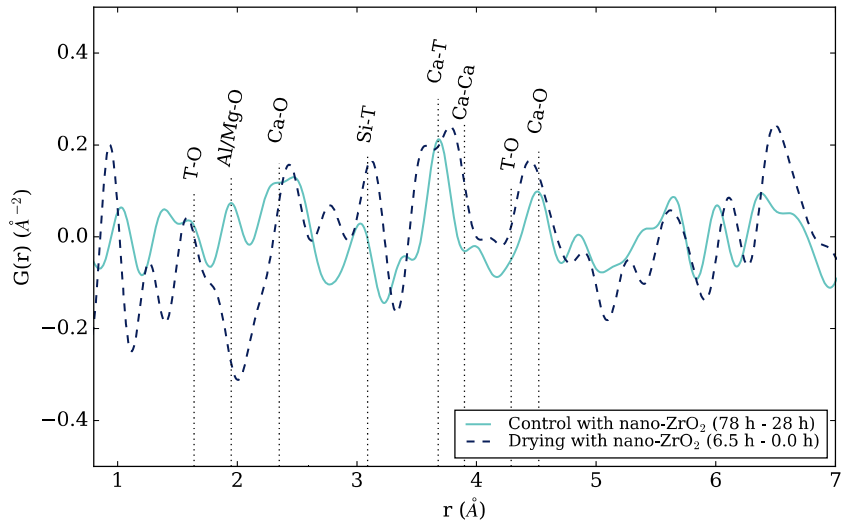

Figure 7. Comparison of the selected difference PDF curves of the silicate-activated slag at 0% RH ("6.5 h – 0.0 h") and in the control environment ("78 h – 28 h") in Figure 6. "0.0 h" indicates the start of drying, where the sample has been cured for 24 hours. "28 h" means that the sample has been cured for 28 hours.

**Kinetics of local structural changes**

To better visualize the changes occurring in the local atomic structure of silicate-activated slag with nano-$ZrO_2$ during drying, and to enable direct comparison with the no nano data given in Figure 2, the PDF peak intensities of the two atom-atom correlations, Si-T and Ca-T, are normalized with respect to T-O peak and plotted against time (Figure 8). Comparing Figures 2 and 8 reveals that the behavior of silicate-activated slag with nanoparticles is in stark contrast with the sample without nanoparticles. The normalized Si-T and Ca-T peak intensities are now seen to increase continuously for both 0% and 43% RH environments. Moreover, the extent of increase is much larger than that seen for the control silicate-activated slag sample during alkali-activation, which, specifically for the (Si-T)÷(T-O) ratio, indicates that the "unconventional" gel that precipitates during drying in these systems has a higher degree of polymerization compared with C-(N)-A-S-H gel.

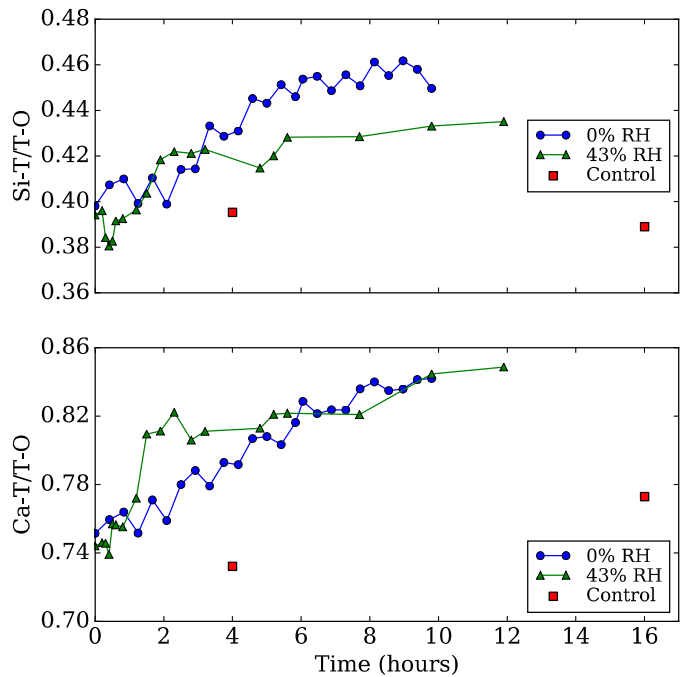

Figure 8. Evolution of normalized Si-T (at ~3.1 Å) and Ca-T (at ~3.65 Å) peak intensities of silicate-activated slag with nano-ZrO$_2$ at 0% RH, 43% RH and in the control (sealed) environments, all normalized with respect to the T-O peak intensity at ~1.65 Å.

Unlike the case of silicate-activated slag without nanoparticles in the 43% RH environment, where the Si-T and Ca-T atom-atom correlations behave similarly to those of the control sample in the sealed environment (Figure 2), the behavior of the Si-T and Ca-T correlations of silicate-activated slag with nano-ZrO$_2$ in the 43% RH environment is closer to that of the 0% RH environment (Figure 8). Specifically, the normalized Ca-T peak intensities of silicate-activated slag with nano-ZrO$_2$ at 0 and 43% RH essentially overlap while the evolution of the normalized Si-T peak intensity of the silicate-activated slag with nanoparticles in the 43% RH environment lies in between that of the silicate-activated slag with nanoparticles in the 0% RH environment and that in the sealed (control) environment. These data show that growth of the "unconventional" C-(N)-A-S-H gel in the silicate-activated slag with nanoparticles occurs even in a moderately dry environment (43% RH).

As seen in Figure 8, the 0% and 43% RH drying experiment was terminated at ~10 and 12 hrs, respectively, determined at the time of measurement by the plateau in the normalized Si-T peak. However, it turns out that the normalized Ca-T peak is still increasing after 10 hrs, indicating that the "unconventional" gel growth mechanism is ongoing. From comparison of Figures 2 and 8 it can be seen that most of the changes occur during the initial 2 hrs for the samples without nanoparticles, while changes are more uniform across the entire measurement time for the samples with nanoparticles (~ 10 hrs for 0% RH case and ~ 12 hrs for 43% RH case). Furthermore, a subsequent weight loss experiment (Figure S10 in Supporting Information) revealed that the evaporation rates of water in both silicate-activated slag with and without nano-$ZrO_2$ are similar, indicating that the continual "unconventional" gel growth in silicate-activated slag with nano-$ZrO_2$ cannot be attributed to the retention of water and therefore and ongoing hydration reaction. Given this, we propose that the continual growth of "unconventional" gel in the silicate-activated slag with nanoparticles is likely associated with the surface reactivity of the nanoparticles, as explained in detail later in this article.

**Strain at the nanoscale**

The impact of nano-$ZrO_2$ on the development of nanoscopic strain during drying is shown in Figure 9, where the associated PDF data are given in Figure S6 in the Supporting Information. These data show that nano-$ZrO_2$ augment the mechanisms occurring at the nanoscale during drying, and therefore mitigate the development of drying-induced shrinkage strains that were previously seen to develop in the sample without nanoparticles (Figure 5). For instance, where a shrinkage strain of 0.0027 developed in the sample without nanoparticles at 0% RH (Figure 5), there is a corresponding expansion strain of 0.0007 for the nano-$ZrO_2$ sample (Figure 9). Although some PDF peaks of the nano-$ZrO_2$ samples (Figure S6) shift considerably (e.g., at 11.1 and 13.3 Å for the 0% RH condition and at 13.3 and 20.3 Å for the 43% RH condition), the majority of them show only minimal change, and some of the exceptionally large values are actually artifacts due to the difficulty in describing the complex peak shape (e.g., at 20.3 Å for the 43% RH condition), as can be seen in Figure S6 in the Supporting Information.

The small expansive strain values in the silicate-activated slag with nano-$ZrO_2$ at 0% RH indicate minimal structural rearrangements in the existing C-(N)-A-S-H gel, possibly due to the reinforcing effect of the "unconventional" gel that precipitates during drying. This interpretation is also supported by the diffraction data (Figure S4b and S5b) where no shift in basal spacing is observed in silicate-activated slag with nano-$ZrO_2$ for both the 0% and 43% RH conditions. Although water is being pulled out the C-(N)-A-S-H gel in the nano-$ZrO_2$ samples during drying (as confirmed by the similar weight loss curves in Figure S10 in the Supporting Information), at 0% RH there is no collapse of the interlayer spacing, and therefore the nanoparticles are providing some sort of reinforcing effect to the C-(N)-A-S-H gel.

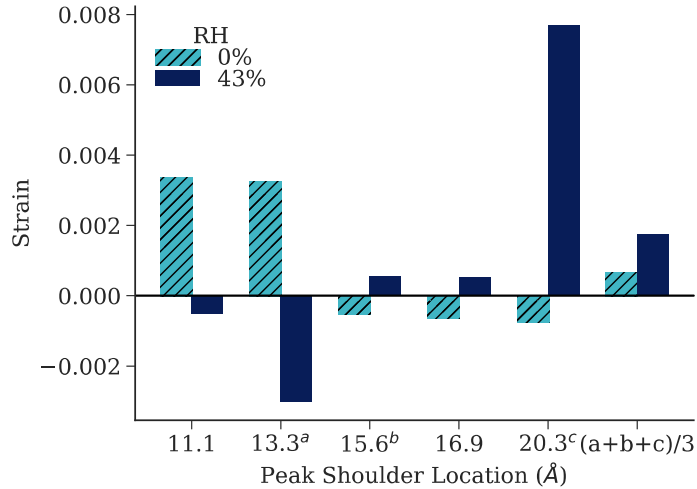

Figure 9. Quantification of the peak shifts in the PDF for silicate-activated slag with nano-ZrO₂ (Figure S6), given as strain values. A positive strain value indicates a shift of the peak shoulder towards larger atomic distances.

### Nano-ZrO₂ catalyzing additional gel growth during drying?

Given the significant impact nano-ZrO₂ has on structural rearrangements/changes that occur during drying, we propose that the surface reactivity of the nanoparticles plays a large role in mitigating drying-induced changes. As water molecules evaporate from silicate-activated slag the ionic concentration of the pore solution will increase. However, for the case of silicate-activated slag without nanoparticles, nucleation of additional C-(N)-A-S-H gel does not occur since local supersaturation levels are not reached. It is known that the pH of the pore solution in the silicate-activated slag paste is around 13.[39] Hence, the surface of the nano-ZrO₂ in contact with the pore solution will be electronegative, since the point of zero charge (pzc) for tetragonal zirconia is around 6.7.[78] These negatively charged surfaces associated with the nanoparticles will lead to adsorption of the calcium cations from the pore solution. A similar phenomenon has been reported to occur on the surface of rutile (titanium oxide).[79] It is likely that at higher ionic concentrations (more aggressive drying conditions), the negatively charged silicate and aluminate ions in the pore solution will also adsorb to the surface of the nano-ZrO₂ (attracted to the positively charged calcium ions), leading to growth of a gel. In fact, growth of C-S-H gel on various nanomaterials has been observed using transmission electron microscopy.[80] PDF data for the alkali-activation of silicate-activated slag containing nano-ZrO₂ show that the nanoparticles remain stable in the alkaline pore solution up to 131 days (Figure S11 in the Supporting Information), and therefore reactive zirconia surface will be available throughout the drying period.

Interestingly, nano-ZrO₂ does not appear to cause the same type of gel growth in silicate-activated slag during the normal alkali-activation reaction, since the PDF data for the silicate-activated slag with (Figure 6b) and without nano-$ZrO_2$ (Figure 1b) as alkali-activation proceeds are essentially the same. Hence, the nanoparticles remain inert during alkali-activation, and it is possible that only once the degree of solvation of the $ZrO_2$ nanoparticles drops below a certain level does the surface reactivity become important. It is also interesting to note that this mechanism does not occur in silicate-activated slag without nano-$ZrO_2$ under drying condition (see Figure 2 and Figure 8), even though there should be an increase in the ionic concentration in these samples. This is due to the fact that the zirconia surface has a higher affinity for binding with ions from the pore solution compared with surfaces that already exist in the silicate-activated slag system (e.g., slag grains, C-(N)-A-S-H gel and hydrotalcite-like phase), such that the growth of the new type of gel in silicate-activated slag with nano-$ZrO_2$ reinforces the nanoscale structure and prevents damage to the original C-(N)-A-S-H gel as water is removed from the sample.

To assess the validity of the proposed mechanism, the interactions between silicate monomeric ions, i.e., $SiO(OH)_3^-$, and C-(N)-A-S-H gel (modeled as 14Å tobermorite) or tetragonal zirconia have been studied using DFT. 14Å tobermorite with a chemical formula of $Ca_5Si_6O_{16}(OH)_2 \cdot 7H_2O$ was used for modeling the C-(N)-A-S-H gel surface. The optimized bulk structure of 14Å tobermorite[81] was cut parallel to the *001* surface. Two different 14Å tobermorite surfaces were studied. For the first case, 14Å tobermorite was cleaved through the interlayer region such that a silica-rich surface was obtained. For the second case, 14Å tobermorite was cleaved through the intralayer Ca-O bonds to obtain a surface that was decorated with calcium atoms. The ground state configurations of the surfaces were relaxed after cutting them from the bulk 14Å tobermorite. Following the surface relaxations, silicate ions were allowed to interact with both of these surfaces, with the most favorable bonding configuration on each surface determined via calculation of the minimum total energy. A similar analysis was carried out for tetragonal zirconia that was cleaved along its *111* surface. Here, tetragonal zirconia was chosen because the PDF of the nano-$ZrO_2$ solution was found to match tetragonal zirconia (see Figure S9 in Supporting Information). Zirconia was cleaved along the *111* surface since it was previously shown to be energetically the most stable surface of tetragonal zirconia.[82] In order to investigate the effect of calcium atoms on the binding of silicate to the surfaces, this process was simulated both with a bare zirconia surface and with a calcium decorated zirconia surface. It should be noted that calcium atoms have been hydrated with five water molecules per unit-cell to begin with for the model C-(N)-A-S-H gel and zirconia.

For each structure, we calculated the binding energy $E_b$ of the silicate ion to the surfaces mentioned above from the expression

$$E_b = E_T[silicate] + E_T[surface] - E_T[silicate + surface]$$

, where

$E_T[silicate], E_T[surface], and E_T[silicate + surface]$

are the total energy of a silicate ion, the surface and the optimized total energy of one silicate ion adsorbed to the surface, respectively. In our notation, $E_b > 0$ indicates that binding of the ion is favorable. All of the total energies were calculated using the same sized slab-based supercell for each structure. Unit-cells for the slabs had dimensions of 12.92 Å × 14.56 Å for 14Å tobermorite and 10.10 Å × 12.25 Å for zirconia in the horizontal ($a \times b$) plane. This corresponds to a separation of 12.92 Å and 12.25 Å between silicates adjacently adsorbed on the surfaces. In the direction perpendicular to the surface (*c*-axis), a vacuum of 20 Å was chosen in order to avoid interaction between repeating unit-cells.

Our results show that the silicate ion binds to both the silica-cleaved C-(N)-A-S-H gel (3.76 eV) and zirconia (3.99 eV) surfaces (Figure 10a), and these binding energies increase in the presence of hydrated calcium atoms (4.76 eV for Ca-rich C-(N)-A-S-H and 8.69 eV for Ca-rich zirconia surface, Figure 10b), especially for the case of zirconia. Hence, calcium ions are seen to play a crucial role during chemisorption of silicate ions on these surfaces. Furthermore, for both cases (bare surfaces and Ca-decorated), silicate ions are attracted much more strongly to the zirconia surface compared to C-(N)-A-S-H. We repeated the calcium decoration of the surfaces without hydration to ensure that the large increase in binding energy for the zirconia surface is not solely attributed to the presence of water molecules. The binding energies of the silicate ions on the dehydrated Ca-rich surfaces (3.85 eV for C-(N)-A-S-H gel, 7.22 eV for zirconia) are seen to be almost as large as those for the hydrated case. Hence, these DFT results strongly support our hypothesis that the presence of zirconia nanoparticles in the silicate-activated slag sample act as templates for additional (and "unconventional") gel growth during drying. For the formation of such "unconventional" gel to occur, two factors are necessary: an increased ionic concentration in the pore solution of the silicate-activated slag, and the presence of highly reactive surfaces such as nano-$ZrO_2$.

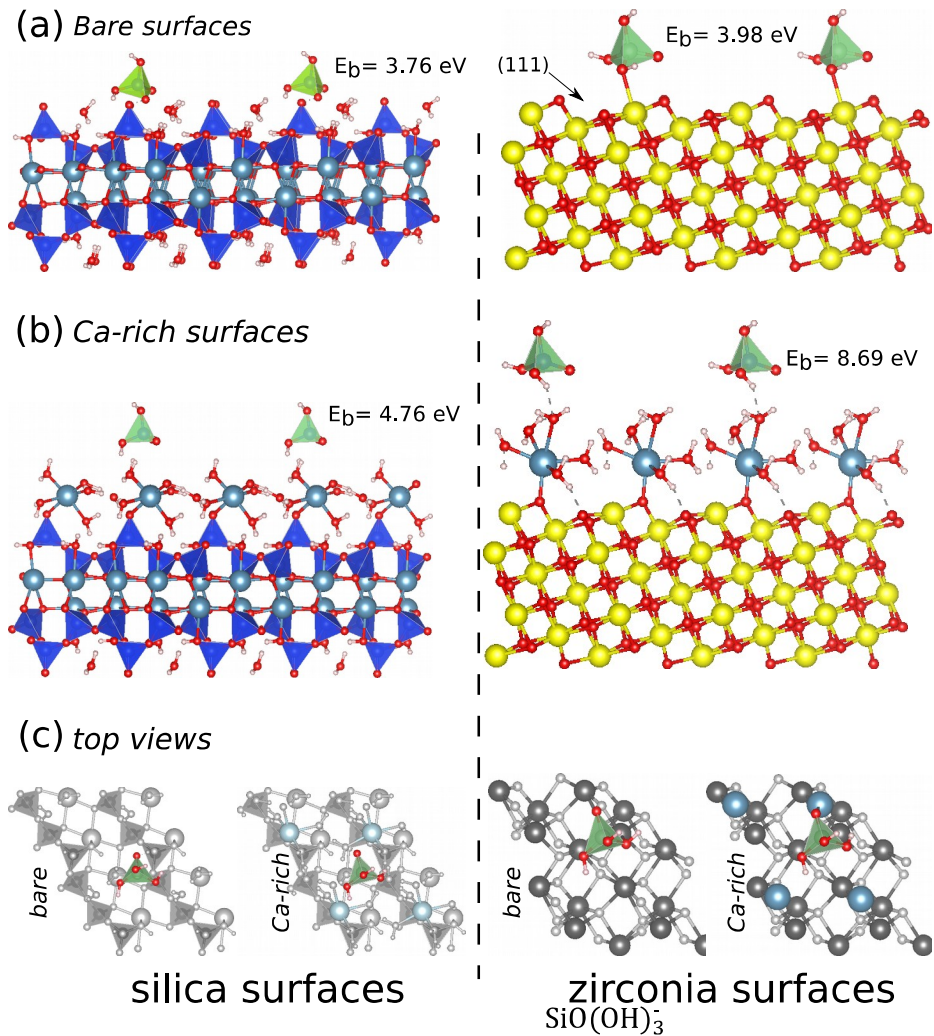

Figure 10. (a) Side views of the optimized geometries of SiO(OH)$_3$ ion adsorbed on 14Å tobermorite (denoted as "silica", model for C-(N)-A-S-H gel) and tetragonal zirconia surfaces, shown on left and the right panels, respectively. (b) Same for Ca-rich hydrated surfaces. (c) Top views showing the growth of a new silica layer on the surfaces. Si, Ca, Zr, O, and H atoms are represented by dark blue, cyan, yellow, red and pink spheres in the ball-and-stick model, respectively. For the top views, the layers that are underneath the top most silica tetrahedrons are shown in gray.

The elucidation of the mechanism reported in this article opens a new pathway to address the long-standing problem of drying-induced microcracking in silicate-activated slag and other materials susceptible to drying-induced cracking. By adding a small amount of zirconia nanoparticles, the silicate-activated slag becomes denser due to additional gel growth during drying, minimizing the extent of nanoscopic shrinkage strain, and therefore potentially making it more resistant to microcracking. However, the nanoparticles should be chosen carefully, as recent studies have shown mixed effects on silicate-activated slag depending on the type of nanoparticles selected.[83,84] The impact of nano-ZrO$_2$ on other cementitious systems (e.g., OPC and alkali-activated metakaolin/fly ash and colloids/ceramics) is also worth investigating, since robust mitigation routes are needed for these crack-prone materials.

## Conclusions

The drying-induced atomic structural changes in sodium-based silicate-activated slag with and without zirconia nanoparticles have been investigated using synchrotron X-ray pair distribution function (PDF) analysis and density functional theory (DFT) calculations. For the case without nanoparticles, the PDF data contain direct information on the development of nanoscopic strain in silicate-activated slag during drying. This strain is caused by two concurrent mechanisms: (1) partial collapse of the interlayer spacing, and (2) slight disintegration of the C-(N)-A-S-H gel leading to breakage of Ca-O-T and Si-O-T linkages and formation of smaller nanosized domains. These findings help bridge the knowledge gap on the nanoscopic shrinkage mechanisms that are contributing to macroscopic length changes in the C-S-H-like systems. Moreover, this investigation has shown that the addition of a small amount of nano-$ZrO_2$ (~ 0.17 wt. %) drastically alters the drying-induced behavior of silicate-activated slag, where an "unconventional" silica-rich gel is seen to precipitate as drying progresses, which provides a reinforcing effect and minimizes the development of nanoscopic strain. Hence, the zirconia surfaces are catalyzing additional gel growth during drying, where an increase in the ionic concentration of the pore solution leads to the precipitation of a highly polymerized silica gel on the surface of the nanoparticles. This proposed mechanism is supported by DFT calculations of the binding energy between a negatively charged silicate monomer and various configurations for C-(N)-A-S-H gel and zirconia surfaces. Irrespective of the hydration state and whether Ca ions are adsorbed to the surfaces, the silicate monomer is found to bind more strongly to zirconia. Therefore, this investigation opens potential pathways for the development of new and novel solutions to the long-standing problem of drying shrinkage in sustainable cements and other porous materials.

## Table of contents

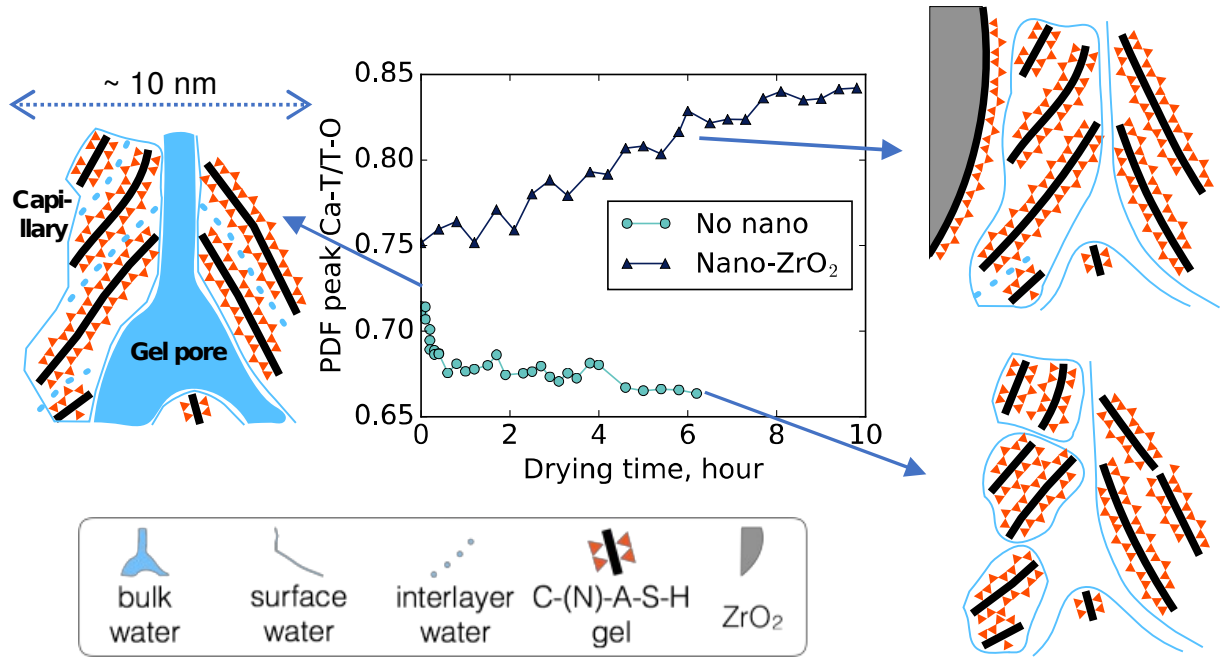